\documentclass[doublecol]{epl2}
\usepackage{amsmath}
\usepackage{amssymb}
\usepackage{amsfonts}
\usepackage{graphicx}
\usepackage{latexsym}
\usepackage{wasysym}
\usepackage{color}

\title{A percolation system with extremely long range connections
and node dilution}

\author{M. L. de Almeida\inst{1}, E. L. Albuquerque\inst{1},
U. L. Fulco\inst{1} \thanks{Corresponding author, e-mail: umbertofulco@gmail.com;
Tel: +-55-84-32153793; Fax: +-55-84-32153791 } \and M. Serva\inst{1,2} }

\shortauthor{M. L. de Almeida, E. L. Albuquerque, U. L. Fulco
and M. Serva}

\institute{

\inst{1} Departamento de Biof\'isica e Farmacologia, Universidade Federal do Rio
Grande do Norte, Natal-RN, Brazil,

\inst{2} Dipartimento di Ingegneria e Scienze dell'Informazione e Matematica,
Universit\`a dell'Aquila, L'Aquila, Italy    \\

}



\abstract{We study the very long-range bond-percolation problem on a linear chain
with both sites and bonds dilution. Very long range means that the probability $p_{ij}$ for a
connection between two occupied sites $i,j$ at a distance $r_{ij}$
decays as a power law, i.e. $p_{ij} = \rho/[r_{ij}^\alpha N^{1-\alpha}]$ when
$ 0 \le \alpha < 1$, and $p_{ij} = \rho/[r_{ij} \ln(N)]$ when $\alpha = 1$.
Site dilution means that the occupancy probability
of a site is $0 < p_s \le 1$.
The behavior of this model results from the competition between long-range
connectivity, which enhances the percolation, and site dilution, which weakens percolation.
The case $\alpha=0$ with $p_s =1 $ is well-known, being the exactly solvable
mean-field model.
The percolation order parameter $P_\infty$ is investigated numerically
for different values of $\alpha$, $p_s$ and $\rho$.
We show that in the ranges $ 0 \le \alpha \le 1$ and $0 < p_s \le 1$
the percolation order parameter $P_\infty$  depends only on
the average connectivity $\gamma$ of sites, which can be explicitly computed
in terms of the three parameters $\alpha$, $p_s$ and $\rho$.}

\begin{document}

\maketitle

\date{\today}

\section{1 - Introduction}

During the last fifty years, percolation theory has brought
new understanding and methods to a broad range of  topics in physics like
materials science, complex networks, surface roughening, epidemiology,
geography, and fire propagation, to cite just a few \cite{stauffer1994,grimmet1999}. 
This theory was first considered for the optimization of masks supplied 
to the miners in the coal pits needing a protection
which could block poisoning materials, while permitting the passage of air. 
In other words, it was needed an
appropriate dosage of porosity of the material which composed the masks
in order to have $connected \; path$ for air and $unconnected \; path$
for poisoning materials.
After that, the theory was applied to the study of movement and filtering of 
fluids through porous materials (the most familiar phenomena probably being
coffee percolation), and its scope has been progressively extended to all
other domains \cite{Bunde,Havlin1,Havlin2}.

Nowadays, percolation is still a very active field of research
and applied to an always increasing number of phenomena in physics 
as, for example, fluid flow in random media\cite{narayan1994}, dielectric breakdown
\cite{sune1990} and  reaction-diffusion processes in two-dimensional 
percolating structures\cite{bianco2013}. 

In the context of percolation theory, a percolation transition is characterized by a set of universal critical exponents, which describe the fractal properties of the percolating medium at large scales and sufficiently close to the transition. The exponents are universal in the sense that they only depend on the type of percolation model and on the space dimension. They are expected not to depend on microscopic details like the lattice structure or whether site or bond percolation is considered \cite{Stanley,Bruce,Yeomans}. 

Percolation models have also been increasingly adopted to many systems in Nature to
understand important features of many chemical, biological, and social phenomena.  These systems differ mainly in their topology: many of them form complex networks,
whose vertices are the elements of the system and whose edges represent the interactions between them. For example, living systems form a huge genetic network,
whose vertices are proteins, while the edges represent the chemical interactions between them \cite{Weng}. Equally complex networks occur also in social science, where the vertices are individuals, organizations or countries, and the edges characterize the social interaction between them \cite{Wass}. Due to their large size and the complexity of the interactions, the topology of these networks is largely unknown or unexplored.

The first percolation model was the well-known Bernoulli (or bond) percolation model, whose generalization  was introduced as the Fortuin-Kasteleyn random cluster model, which has many connections with the Ising  and  Potts models in the language of percolation theory. \cite{Fortuin1,Fortuin2,Fortuin3}. Furthermore, the effect of connectivity on biodiversity can be supported by percolation theory hypothesizing the existence of a connectivity threshold, whose measurement is considered as a practical means to facilitate biological fluxes and optimize species colonization possibilities in designing a policy of conservation biology. Recent advances in this field points to  universal laws and offer a new conceptual framework that could potentially revolutionize our view of biology and disease pathologies in the twenty-first century \cite{Barabasi}. Applications of complex networks, based largely on graph theory, have been rapidly translated to characterize  brain network organization as well \cite{Bull}.

The effect of long-range connections on percolation is of fundamental interest, 
since they give rise to a variety of new interesting dynamical and thermodynamical 
phenomena. In view of that,  long-range models have  been  intensively studied 
in recent times in different context 
\cite{rego1999,silva2002,koval2012,zhang2013,schrenk2013}.
The phenomenology becomes very interesting when long range connections appears
together with site dilution.  In this case, the presence of competitions between 
long-range connectivity  enhance percolation and site dilution, leading to a weaker percolation effect
\cite{fulco2003,fulco2003b,albuquerque2005}.

In this work, we want to investigate
the very long-range bond-percolation problem on a linear chain
with  both sites and bonds dilution. 
Very long range means that the probability $p_{ij}$ for a
connection between two occupied sites $i,j$ at a distance $r_{ij}$
decays as a power law, i.e. $p_{ij} = \rho/[r_{ij}^\alpha N^{1-\alpha}]$ when
$ 0 \le \alpha < 1$, and $p_{ij} = \rho/[r_{ij} \ln(N)]$ when $\alpha = 1$. 
Site dilution means that the occupancy probability of a site is $0 < p_s \le 1$.
Notice that for this very long range models
it is necessary to assume that the probability of 
connection decays as $1/ N^{1-\alpha}$ (or $1/\ln(N)$) in 
order to obtain the correct thermodynamic limit.

The case $\alpha=0$, with $p_s =1 $, is well-known, being the exactly solvable
mean-field model, while the case $\alpha=0$ with $p_s <1 $ 
is its almost trivial extension.
In the other regions, the percolation order parameter $P_\infty$ is investigated numerically
for different values of $\alpha$, $p_s$ and $\rho$.
Intuitively, one expects the percolation order parameter $P_\infty$ 
be reduced by the inclusion of diluted sites
\cite{fulco2003,fulco2003b}.
Indeed, we will show not only that this is true, but we also  show
that in all range $ 0 \le \alpha \le 1$, $0 < p_s \le 1$
the percolation order parameter $P_\infty$  depends only on
the average connectivity $\gamma$ of sites, which can be explicitly computed
in terms of the three parameters $\alpha$, $p_s$ and $\rho$.
Besides, the connectivity reduces when dilution increases.

From a technical point of view, we will show that for values of $\alpha$ 
either between $0.2$ and $0.8$, or for $\alpha=1$, considering  different values of $p_s$, all curves collapse.

The paper is organized as follows:
In section 2 we consider the simple case $\alpha=0$
in the absence and in the presence of dilution. Sections 3 and 4 discuss the 
cases $0<\alpha<1$ and 
$\alpha=1$ respectively. Finally, our conclusions are depicted in section 5.

\section{2 - Mean-field ($\alpha$=0)}

The percolation order parameter $P_\infty$ is defined
as the fraction of sites of the system that belong to the
infinite cluster. Obviously, $P_\infty$ attains its maximum value
($P_\infty = 1$) when all the sites of the system appear inside the
infinite cluster, whereas $P_\infty = 0$ below a certain threshold,
when it is not possible to produce an infinite cluster.

A particularly simple model is the mean-field, which corresponds
to $\alpha=0$.  We describe below this almost trivial case,
first when only bonds are diluted, and afterwards  considering dilution for both 
 sites and bonds. 

\subsection{a - Mean-field (bond diluted)}

\begin{figure}[!ht]
 \includegraphics[width=3.4truein,height=3.4truein,angle=0]{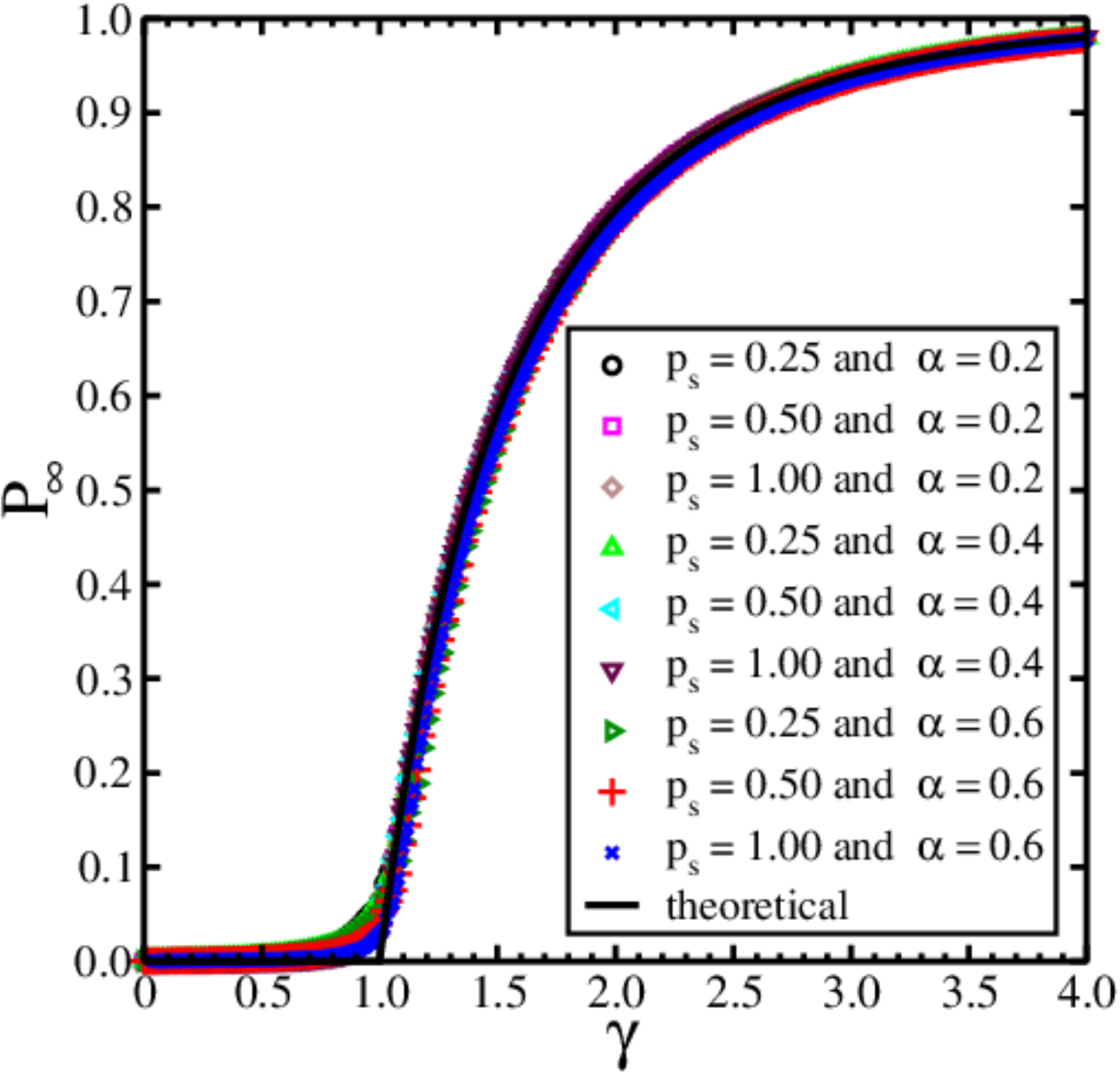}
\caption{\small Order parameter $P_\infty$ versus the control parameter
$\gamma (\rho, \alpha, p_s)$. For a given choice of the
parameters $\alpha$ and $p_s$, $\gamma$ only depends on $\rho$.
The figure shows a complete collapse, i.e. the shape is the same
for any  choice of $\alpha$ and $p_s$, and coincides
with the well-known mean-field result.}
\end{figure}

In mean-field bond diluted model ($\alpha=0, \, p_s =1$), 
one assumes  that there are $N$ nodes.
Any pair of nodes is connected (closed bond) with probability
$\rho/N$ and unconnected (open bond) with probability $1-\rho/N$.

The average connectivity $\gamma$ of a given node (the average number of
connections of a node to the remaining $N-1$ nodes) is  given by

\begin{equation}
\gamma = \frac{\rho}{N}\,
(N-1) \simeq \rho.
\end{equation}
This number is simply obtained by multiplying the
number $N-1$ of remaining nodes with the probability
that a bond is closed.

Let us call $P_\infty $ the fraction of nodes in
the giant component (number of nodes in the giant component divided by
the total number of nodes $N$),
which can also be seen as the probability that a node
belongs to the giant component itself.
The order parameter $P_\infty $
satisfies the self-consistency equation
(see, for example, \cite{serva2010, serva2011})

\begin{equation}
\exp(-\gamma P_\infty) = 1-P_\infty,
\end{equation}
which gives $P_\infty(\gamma)$.
The critical value of the control parameter is $\gamma_c = 1$,
as it is depicted in Fig.\ 1.

\subsection{b -Mean-field (bond and node diluted)}

This model has a distribution of bonds as the previous model but
node dilution is introduced.
It is assumed that a node is occupied (active node) with probability
$p_s$, and empty (inactive node) with probability $1-p_s$.
The number of active nodes $N_s$ is about $p_s N$.

The average connectivity $\gamma$ of a given active node is

\begin{equation}
\gamma = \frac{\rho}{N}\,
(N_s-1) \simeq \rho \, p_s.
\end{equation}
This number is simply obtained by multiplying the
number $ N_s -1 \simeq p_s \, N$ of remaining active nodes with
the probability $\rho /N$ that a bond is closed.

It is easy to show that  the size  $P_\infty(\gamma)$ of the giant component
is still the function of the average connectivity given by eq.\ (2).
In fact, it is sufficient to remark that the size of the giant component
is, by definition, the number of active nodes in the giant component
divided by the total number of active nodes.
Therefore, it is enough to consider a system composed only by active nodes
(whose number $N_s$ is about $p_s \, N$) which are connected (active bond) with
a probability $\rho/N \simeq\rho p_s /N_s$.
In this way, we are re-conduced to the previous model with the difference that
$\rho$ and $N$ are replaced by $\rho \, p_s$  and $N_s$ in eq.\ (1).
Observe that the average connectivity was $\gamma = \rho $ in previous model,
while now $\gamma = \rho \, p_s$. So, eq.\
(2) must hold also for the present model with $\gamma$ given by (3).

Notice that $\gamma$ increases linearly with $p_s$, while
 $P_\infty$ is a non-decreasing function of $\gamma$.
Therefore, dilution decreases the value of $P_\infty$ as expected.

\section{3 - Power law model (0$<\alpha<$1)}

\subsection{a - Definitions}

Here we consider a one-dimensional (periodic chain)
problem where nodes are active
according to a given rate $p_s$. Two nodes are connected (closed bonds),
at a power law probability depending on their distance.

Assuming that $i$ is the position of a node on the chain,
periodic boundary conditions implies that the node $i$
coincides with node $i+N$.
Furthermore, given the periodic boundary
conditions, the distance  $r_{ij}$ between two nodes $i$ and $j$
is $r_{ij}=|i-j|$ if $1 \le |i-j| \le N/2$ and
$r_{ij}=N-|i-j|$ if $N/2 <|i-j| < N$.

Therefore, we assume that two active nodes $i$ and $j$ of the chain
are connected by a closed bond
depending upon their distance $r_{ij}$ according to the probability
$P(r_{ij})$ which obeys a power-law, i.e.

\begin{equation}
p_{ij}=\frac{\rho}{(r_{ij})^{\alpha}\, N^{1-\alpha}},
\end{equation}
with $ 0 \le\alpha < 1$.
According to the above prescription, nearest neighbors nodes (when both are active)
are connected with probability $\rho/N^{1-\alpha}$.

The exponent $ 0 \le\alpha < 1$  controls the range of network interaction.
When $\alpha = 0$, the model reduces to the
model described in subsection $b$ of section 2.

\subsection{b - Solution}

\begin{figure}[!ht]
 \includegraphics[width=3.4truein,height=3.4truein,angle=0]{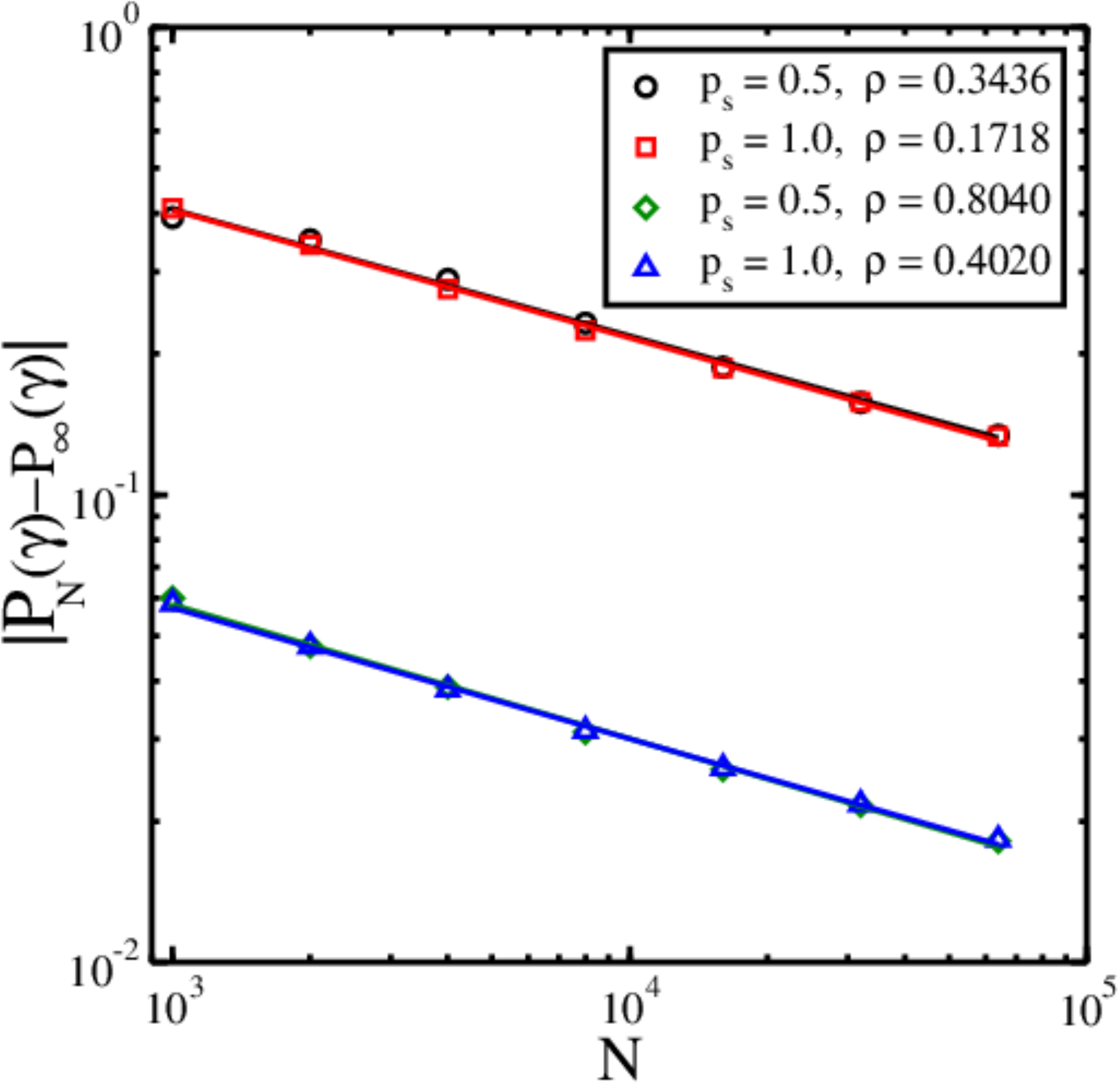}
\caption{\small
The function $P_N (\lambda)-P_\infty(\lambda)$  plotted against $N$ in a log-log scale
for the case $\alpha=0.8$.}
\end{figure}

The probability that a given active node at position $i$
is connected to another active node at position $j$ at distance $r$ is

\begin{equation}
\frac{\rho \, p_s}{r^\alpha \, N^{1-\alpha}}.
\end{equation}

Given the periodic boundary conditions, there are two nodes at any
given distance $1 \le r \le N/2$.
Therefore, the average connectivity $\gamma= \gamma (\rho, \alpha, p_s)$
is obtained by the sum

\begin{equation}
\gamma = \sum_{r=1}^{N/2}\, 2\, \frac{\rho \, p_s}{r^\alpha \, N^{1-\alpha}}
\simeq \frac{2^\alpha}{1-\alpha} \, \rho \, p_s,
\end{equation}
where we have neglected terms which vanish in the thermodynamical limit.
Remark that (6) reduces to (3) when $\alpha \to 0$, and that the average distance $d$ 
between two active connected nodes of the system
has the the size of a finite fraction of the the system size.
In fact:

\begin{equation}
d =  \frac{1}{\gamma}
\sum_{r=1}^{N/2} \, 2 \, r \, \frac{\rho \, p_s}{r^\alpha \, N^{1-\alpha}}
\simeq  \, \frac{1 -\alpha}{2 \, (2-\alpha)} \, N.
\end{equation}
This fact should imply mean-field properties for the system.
Given that $P_\infty$ is the ratio between
the number of active nodes in the giant component and the total number
of active nodes, $P_\infty(\gamma)$ should be still given by eq.\ (2),
provided that $\gamma $ is given by (6), which is the aim  of the numerical work.

In practice, for any value of $\alpha$, $\rho$ and $p_s$ one should compute
numerically $P_\infty$ and plot against $\gamma=\gamma(\alpha,\rho,p_s)$
given by (6).
Once obtained a curve, one should compare it with the mean-field curve
given by eq.\ (1). In Fig.\ 1 we show that, indeed,
the shape of $P_\infty(\lambda)$ is the same
independently on the numerical parameters, and coincides
with the well known mean-field result.

For a given choice of the parameters $\alpha$ and $p_s$,
$\gamma$ only depends on $\rho$. Therefore,
we have considered various values of $p_s$ and $\alpha$
and plotted $P_\infty(\lambda)$  with respect to $\lambda$.
For all cases we have considered a system of $N=10,000$ sites,
and we have obtained $P_\infty$ as an average over $500$ different
independent realizations of the network.

Fig.\ 1 only shows results up to $\alpha= 0.6$ because for larger
values of $\alpha$, a size $N=10,000$  is not enough
to reach the thermodynamical limit. Nevertheless, we are able to show,
by a scaling analysis,
that the mean-field value is anyway reached in the $N \to \infty$ limit.
This can be seen in Fig.\ 2 where the difference
$P_N (\lambda)-P_\infty(\lambda)$ is plotted against $N$ in a log-log scale
for the case $\alpha=0.8$.
The function $P_\infty(\lambda)$ is the value calculated analytically, while
$P_N (\lambda)$ is the value obtained by a simulation of a network of size $N$.
We observe that the difference $P_N (\lambda)-P_\infty(\lambda)$
converges  as a power law to zero, confirming our data collapse
for various values of the parameters $\rho$ and $p_s$.
In particular, the upper (lower) data of Fig.\ 2 correspond to
two different choices of the parameters, both considering $\gamma$ = 1.5 (3.5).
In all cases, the  values of $P_N (\lambda)$ are obtained as an average
over $500$ different  independent realizations of the network, with $N$ ranging from 1,000 to 64,000.

\begin{figure}[!ht]
 \includegraphics[width=3.4truein,height=3.4truein,angle=0]{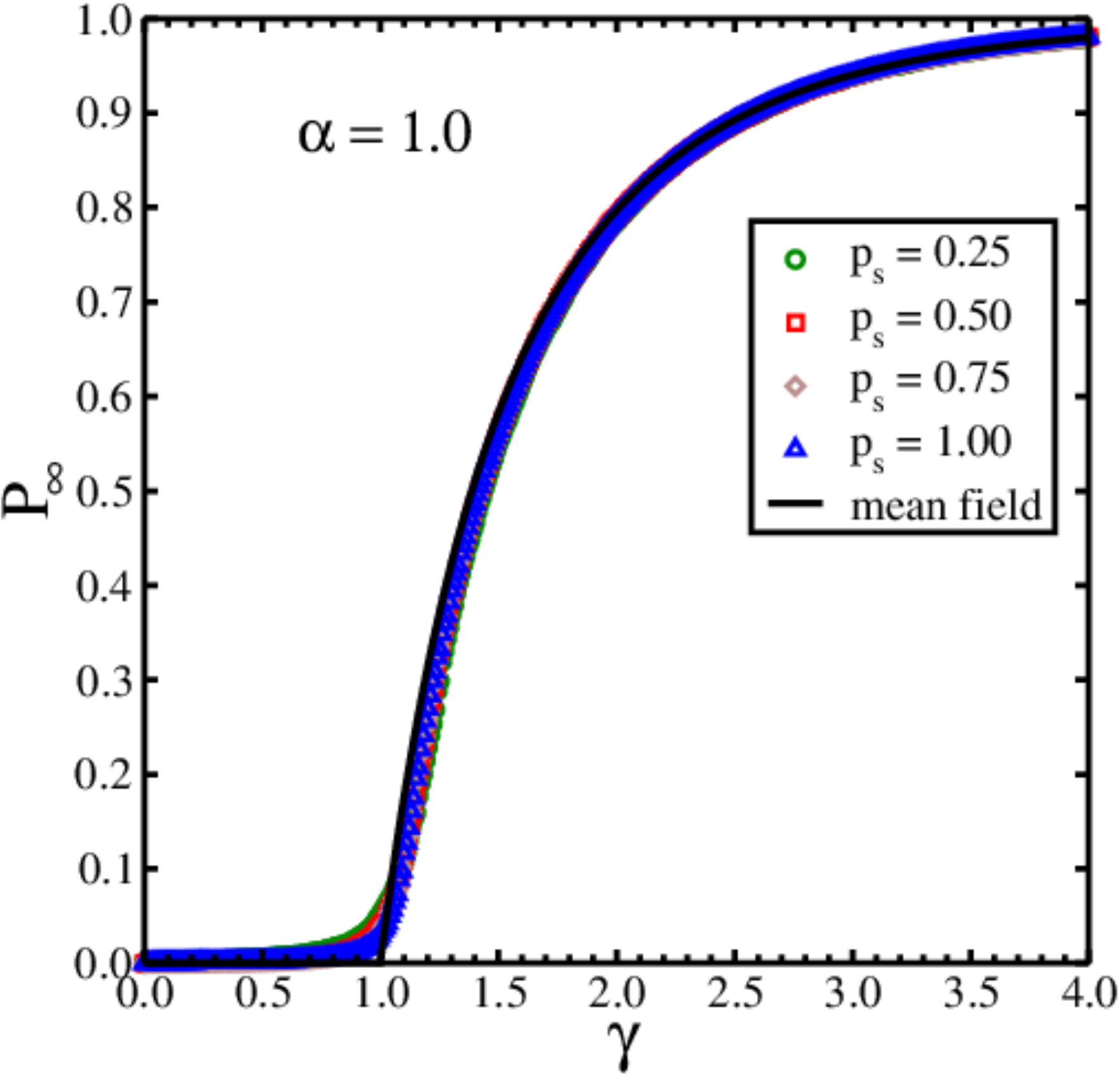}
\caption{\small Order parameter $P_\infty$ versus the control parameter
$\gamma (\rho, p_s)$ for $\alpha=1$. For a given choice of the
parameter $p_s$, $\gamma$, as in Fig.\ 1, only depends on $\rho$.
The figure shows again a complete collapse.}
\end{figure}

\section{4 - The case $\alpha$ = 1}

When $\alpha = 1$ everything goes
as before but scaling $1/\ln(N)$ replaces scaling $1/N^{1-\alpha}$
in eq.\ (4), i.e.:

\begin{equation}
p_{ij}=\frac{\rho}{r_{ij}\, \ln(N)}.
\end{equation}
Then, the average connectivity is

\begin{equation}
\gamma = \sum_{r=1}^{N/2}\, 2\, \frac{\rho \, p_s}{r\, \ln(N)} \simeq
2\, \rho \, p_s
\end{equation}

We remark that now, at variance with the case $0\le \alpha < 1 $,
the average distance $d$ between two active
connected nodes of the system is not of the order of the size of the system.
In fact:

\begin{equation}
d =  \frac{1}{\gamma}
\sum_{r=1}^{N/2} \, 2 \, \frac{\rho \, p_s}{\ln(N)}
\simeq  \, \frac{1}{2} \frac{N}{\ln(N)}.
\end{equation}
Nevertheless, the system still has mean-field properties.

We have checked  numerically that
$P_\infty(\gamma)$ given by eq.\ (2) should  still be valid,
provided  $\gamma $ is given by (9), as it is shown
 in Fig.\ 3, where $P_\infty(\gamma)$ is plotted
against $\gamma$.
Considering $\alpha=1$, for a given choice of the parameters $p_s$,
$\gamma$ only depends on $\rho$.
We have taken into account various values of $p_s$ and, for all cases, a system
with $N=10,000$ sites. $P_\infty$ is obtained as an average over $500$
different independent realizations of the network.

\section{5 - Discussion}

We have shown that very long-range bond-percolation ($ 0 \le \alpha \le1$),
with  both sites and bonds dilution on a linear chain behaves 
as in the mean-field theory if $P_\infty$
is expressed in terms of the average connectivity $\gamma$ of a site.
In other words, all data collapse on the same universal curve if 
$P_\infty$ is plotted against $\gamma(\alpha,\rho,p_s)$.

Noticeably, collapse is absent when the parameter $\alpha$ is larger 
then unity \cite{rego1999,fulco2003,fulco2003b}.
In this case, in fact, connectivity is not an exhaustive description of the
topology of the system, as it can be easily understood.
If $\alpha>1$ one can assume $p_{ij}=\rho/r_{ij}^\alpha$ and then compute 
the average connectivity 

\begin{equation}
\gamma=\sum_{r=1}^{N/2}\, 2\, \rho \, p_s/r^\alpha
\simeq 2\, \rho \, p_s \zeta(\alpha), 
\end{equation}
where $\zeta(\alpha)=\sum_{r=1}^{\infty}\, r^{-\alpha}$
is the zeta Riemann function.
In this $\alpha >1$ case, at variance with the case $0 \le \alpha \le 1$,
the probability of connections to sites at a distance of order unity
remains finite in the thermodynamic limit.

This implies that fluctuations in connections
with few close sites and bonds, may prevent  the emergence of a giant 
component even when $\gamma>1$.

To clarify this point consider the simple problem of percolation on a linear chain
occupancy probability $p_s$ and 
with connections only between nearest-neighbor sites.
This case corresponds to the limit $\alpha \to \infty$ where 
$p_{ij}=\rho/r_{ij}^\alpha$ equals 1 if $r_{ij}=1$,  vanishing otherwise.
In this case, whenever $\rho<1$ or $p_s<1$  percolations is forbidden.
In other words  $P_\infty =1$ for $p_s\rho = 1$ and $P_\infty=0$ for $ p_s \rho<1$.
Since the average connectivity is $\gamma=2\, \rho \, p_s $, since $\lim_{\alpha \to \infty}\zeta(\alpha)=1$,
one can also state  $P_\infty =1$ for $\gamma = 2$
(which is the maximum possible value for $\gamma$)
and $P_\infty=0$ for $\gamma<2$.
This behavior is at very variance with those described in this paper for the 
very long range model.

We would like to finally stress that while
we are always able to  explicitly compute $\gamma$ in terms of 
$\alpha$, $p_s$ and $\rho$ for all possible range of these three parameters,
the case $\alpha=0$ is the only one which we are able to  
completely treat analytically.
Our conclusions concerning the region $ 0 < \alpha \le 1$
are mainly based on numerical simulations.
Our results are very precise and hopefully correct, although
 a rigorous mathematical proof of our conclusions still remains open.

\begin{acknowledgments}

This work was partially supported by the Brazilian Research Agencies
CAPES (Rede NanoBioTec and PNPD), CNPq (Procad-Casadinho) and 
FAPERN/CNPq (PRONEM). 
M.S. was partially supported by PRIN 2009 protocollo n. 2009TA2595.02.

\end{acknowledgments}


\begin{thebibliography}{00}


\bibitem{stauffer1994} D. Stauffer and A. Aharony,
{\it Introduction to Percolation Theory},
Taylor and Francis, London, 1994.

\bibitem{grimmet1999} G. Grimmet, 
{\it Percolation}, 
2nd Edition, Springer-Verlag, Berlim, 1999.

\bibitem{Bunde} A. Bunde and S. Havlin (eds): {\it Fractals and Disordered Systems}, 2nd Edition, Springer-Verlag, Berlin, 1996

\bibitem{Havlin1} D. ben Avraham and S. Havlin, {\it Diffusion and Reactions in Fractals and Disordered Systems}, Cambridge University
Press, Cambridge, 2000. 

\bibitem{Havlin2} R. Cohen and S. Havlin, {\it Complex Networks: Structure, Robustness and Function}, Oxford University
Press, Oxford, 2010. 

\bibitem{narayan1994} O. Narayan and D.S. Fisher, 
{\it Nonlinear fluid flow in random media: Critical phenomena near threshold},
Phys. Rev. B {\bf 49}, (1994) 9469.

\bibitem{sune1990} J. Sune, I. Placencia, N. Barniol, E. Farres, F. Martin and X. Aymerich, 
{\it On the breakdown statistics of very thin SiO$_2$-films}, 
Thin Solid Films {\bf 185}, (1990) 347.

\bibitem{bianco2013}
F. Bianco, F. Chibbaro, D. Vergni and A. Vulpiani, 
{\it Reaction spreading on percolating clusters}
Phys. Rev. E {\bf 87}, 062811 (2013).

\bibitem{Stanley} H.E. Stanley, {\it Introduction to Phase Transitions and Critical Phenomena}, Oxford University Press, Oxford, 1971.

\bibitem{Bruce} A.D. Bruce and R.A. Cowley, {\it Structural Phase Transitions}, Taylor and Francis, London, 1981.

\bibitem{Yeomans} J.M. Yeomans, {\it Statistical Mechanics of Phase Transitions}, Oxford University Press, Oxford, 1992.

\bibitem{Weng} G. Weng, U. S. Bhalla and R. Iyengar, {\it Complexity in biological signaling systems}, Science {\bf 284}, (1999) 92. 

\bibitem{Wass} S. Wasserman and K. Faust, {\it Social Network Analysis}, Cambridge University Press, Cambridge, 1994.

\bibitem{Fortuin1} C.M. Fortuin and P.W. Kasteleyn, {\it Random-cluster model: I. Introduction and relation to other models}, Physica  {\bf 57}, (1972) 536.

\bibitem{Fortuin2} C.M. Fortuin, {\it Random-cluster model: II. Percolation model}, Physica  {\bf 58}, (1972) 393.

\bibitem{Fortuin3} C.M. Fortuin, {\it Random-cluster model: III.  Simple random-cluster model}, Physica  {\bf 59}, (1972) 545.

\bibitem{Barabasi} A-L Barab\'{a}si and Z.N. Oltvai {\it Network biology: understanding the cell's functional organization}, Nature Reviews Genetics {\bf 5} (2009) 101. 

\bibitem{Bull} E. Bullmore and O. Sporns {\it Complex brain networks: graph theoretical analysis of structural and functional systems}, Nature Reviews Neuroscience {\bf 10} (2004) 186. 

\bibitem{rego1999} H.H.A. Rego, L.S. Lucena, L.R. da Silva and C. Tsallis, 
{\it Crossover from extensive to nonextensive behavior driven by long-range $d=1$ bond percolation}, Physica A  {\bf 266}, (1999) 42.

\bibitem{silva2002} C.R. da Silva, M.L. Lyra and G.M. Viswanathan, 
{\it Largest and second largest cluster statistics at the percolation 
threshold of hypercubic lattices}, 
Phys. Rev. E {\bf 66}, (2002) 056197.

\bibitem{koval2012} V. Koval, R. Meester and P. Trapman,
{\it Long-range percolation on the hierarchical lattice}
Electron. J. Probab. {\bf 17}, (2012) 1.

\bibitem{schrenk2013} K.J. Schrenk, N. Pose, J.J. Kranz et al., 
{\it Percolation with long-range correlated disorder}
Phys. Rev. E {\bf 88}, (2013) 052102.

\bibitem{zhang2013} Z. Zhang and L. Zhang, 
{\it Scaling limits for one-dimensional long-range percolation: Using the corrector method},
Stat. Prob. Lett., {\bf 83}, (2013) 2459.

\bibitem{fulco2003} U.L. Fulco, L.R. da Silva, F.D. Nobre, H.H.A. Rego and L.S. Lucena, 
{\it Effects of site dilution on the one-dimensional long-range bond-percolation problem}, 
Phys. Lett. A {\bf 312}, (2003) 331.

\bibitem{fulco2003b} U.L. Fulco, L.R. da Silva, F.D. Nobre and L.S. Lucena, 
{\it Competing Long-Range Bonds and Site Dilution in the One-Dimensional Bond-Percolation Problem}, Braz. J. Phys. {\bf 33}, (2003) 645.

\bibitem{albuquerque2005} S.S. Albuquerque, F.A.B.F. de Moura, M.L. Lyra and A.J.F. de Souza, 
{\it Fractality of largest clusters and the percolation transition in power-law diluted chains}, Phys. Rev. E {\bf 72}, (2005) 016116.

\bibitem{serva2010} M. Serva, 
{\it Magnetization densities as replica parameters: the dilute ferromagnet},
Physica A {\bf 389}, (2010) 2700.

\bibitem{serva2011} M. Serva, 
{\it Exact and approximate solutions for the dilute Ising model},
Physica A {\bf 390}, (2011) 2443.




\end{thebibliography}
\end{document}